\documentclass[useAMS,usenatbib]{mn2e}

\usepackage{amssymb}
\usepackage{graphicx}

\newcommand{\Ms}{\ensuremath{M_{\odot}}}
\newcommand{\eg}{{\it e.g.}}
\newcommand{\cf}{{\it c.f. }}
\newcommand{\ie}{{\it i.e.}}
\newcommand{\viz}{{\it viz.}}
\newcommand{\imfmax}{\ensuremath{m_{\rm max\ast}}}
\newcommand{\tsc}{\ensuremath{\tau_{\rm SC}}}

\title[Super-canonical stars in R136]{The emergence of super-canonical stars in R136-type star-burst clusters}
\author[Banerjee, Kroupa and Oh]{Sambaran Banerjee$^{1}$\thanks{E-mail: sambaran@astro.uni-bonn.de (SB)},
Pavel Kroupa$^{1}$\thanks{E-mail: pavel@astro.uni-bonn.de (PK)} and Seungkyung Oh$^{1}$\thanks{E-mail: skoh@astro.uni-bonn.de (SKOH)}\\
$^{1}$Argelander-Institut f\"ur Astronomie, Auf dem H\"ugel 71, D-53121, Bonn, Germany}
\begin{document}

\date{Accepted {~~~~~~~~~~~~~~~~~~~~}. Received {~~~~~~~~~~~~~~~~~~~~}; in original form {~~~~~~~~~~~~~~~~~~~~}.}

\pagerange{\pageref{firstpage}--\pageref{lastpage}} \pubyear{0000}

\maketitle

\label{firstpage}

\begin{abstract}
Among the most remarkable features of the stellar population of R136, the central, young, massive star cluster
in the 30 Doradus complex of the Large Magellanic Cloud, are the single stars
whose masses substantially exceed the canonical stellar upper mass limit of $150\Ms$. A recent study by us,
\viz, that of Banerjee, Kroupa \& Oh (2012; hereafter Paper I) indicates that such ``super-canonical'' (hereafter
SC) stars can be formed out of a dense stellar population with a canonical initial mass function (IMF)
through dynamically induced mergers of the most massive binaries. The above study consists of
realistic N-body computations of fully mass-segregated star clusters mimicking R136 in which all the massive stars
are in primordial binaries. In the present work, we study the formation of SC stars in the computed R136 models
of Paper I in detail. Taking into consideration that extraneous SC stars form in the computed models
of Paper I due to the primordial binaries' initial eccentricities, we compute additional models where all
primordial binaries are initially circular. We also take into account 
the evolution of the mass of the SC stars and the resulting lifetime in their SC phase using detailed
stellar evolutionary models over the SC mass range that incorporate updated treatments of the stellar winds.
In all these computations, we find that SC stars begin to
form via dynamical mergers of massive binaries from $\approx 1$ Myr cluster age. We obtain SC
stars with initial masses up to $\approx250\Ms$ from these computations. Multiple SC stars are found to remain
bound to the cluster simultaneously within a SC-lifetime.
However, we also note that SC stars can be formed at runaway velocities
which escape the cluster at birth. These properties of the dynamically formed SC stars, as obtained from our computations,
are consistent with the observed SC stellar population in R136. In fact, the
evolutionary models of SC stars imply that had they formed primordially along with the rest of the R136 cluster,
\ie, violating the canonical upper limit, they would have evolved below the canonical $150\Ms$ limit by
$\approx 3$ Myr, the likely age of R136. Thus
according to the new stellar evolutionary models, primordially-formed SC stars should not be observable
at the present time in R136. This strongly supports the dynamical formation scenario
of the observed SC stars in R136.
\end{abstract}


\section{Introduction}\label{intro}

The study of the functional form of the number distribution of stars in galaxies, with which they are born,
or the stellar initial mass function (IMF) has
always been of fundamental importance \citep{bast2010,pk2011}. The high-mass end of the stellar IMF is in particular focus
due to the feedback it gives to the star-forming gas in the form of radiation pressure, kinetic energy from the stellar
wind and supernova ejecta and as well the chemical enrichment from the latter source. An upper limit of $\approx 60\Ms$
to the mass of a star could be set by the ``Eddington limit'' \citep{edd26} or the point of balance of gravity by the radiation
pressure of a star. However, in practice stellar masses can easily exceed the Eddington limit since massive
stars are not fully radiative but contain convective cores \citep{kw90}. The next upper limit is set by the possibility
of the destruction of a star due to thermal pulsations \citep{sh59,bm94}. \citet{stroh92} determined this limit to be at
$\imfmax\approx 120-150\Ms$ for [Fe/H]$\approx 0$ and $\imfmax\approx 90\Ms$ for [Fe/H]$\approx -1$.

Difficulties also arise if one considers the growth of a proto-star via gas accretion which is crucial for
massive star formation. Here one again encounters the $\approx 60\Ms$ Eddington limit.
Stellar formation models lead to a mass limit near
$40-100\Ms$ imposed by feedback on a spherical accretion envelope \citep{kah74,wolfc87}.
Some observations suggest that proto-stars may be accreting material in disks rather than spheres
(\eg, \citealt{chi2004}) in which case it may be possible to overcome the radiation pressure at the equator of the proto-star.
Studies on the formation of massive stars through disk-accretion with high accretion rates,
thereby allowing the radiation to escape preferentially along the poles (\eg, \citealt{jija96})
indeed allow formation of stars with larger masses.

The feedback-induced mass limit can also be avoided if massive stars can form through mergers \citep{bonn98a,zy2007}.
In this scenario, massive stars form via coalescence of intermediate-mass proto-stars in the cores
of dense stellar clusters that have undergone core-contraction due to rapid accretion of
gas with low specific angular momentum. The high central density required for the mergers ($10^8\Ms$ pc$^{-3}$) is still
difficult to achieve but it should be noted that an observable young cluster is necessarily
disposed of a substantial fraction of its natal cloud and is thus likely to be always observed in an expanding
and hence diluted phase. Recently-set limits on the radii of embedded clusters indeed suggest them to form
very compact \citep{mrk2012}. In this context, it is worthwhile to note that in the computations discussed
in the following sections we do get binary coalescence events even in moderately dense stellar clusters. This happens
due to the presence of a substantial number of binaries in the cluster core which have much larger encounter
cross sections than single stars due to their bigger geometrical sizes.

While the existence of an upper limit of the stellar mass is rather obvious from theory since several decades,
such an upper limit has been established from observations only recently. There have been some earlier indications of
the presence of a cut-off near $\imfmax\approx 150\Ms$ from observations of young massive clusters, particularly
in R136 \citep{mhunt98,massi2003} in the LMC and in Arches cluster \citep{fig2003} close to our Galactic center,
but these results were not sufficiently
convincing in the sense that no statistical significance were attached to them. The observed upper limit was considered
to be a limitation due to sampling rather than a true limit \citep{massi2003}.
\citet{elm2000} also noted that random sampling from
an unlimited IMF for all star-forming regions in the Milky Way would lead to the
prediction of stars with masses $\gtrsim 1000\Ms$, unless there is a sharp down-turn in
the IMF beyond several $100\Ms$.

\begin{figure*}
\centering
\includegraphics[height=6.0cm,angle=0]{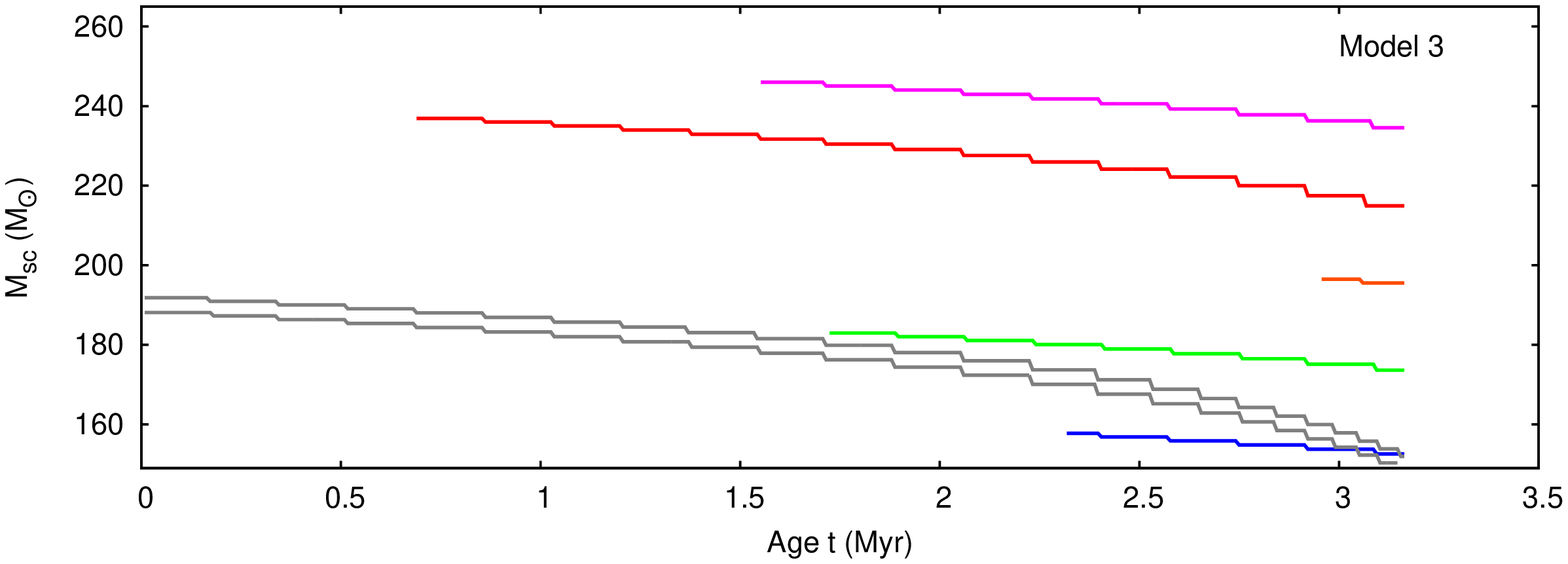}\\
\vspace{-0.5cm}
\includegraphics[height=6.0cm,angle=0]{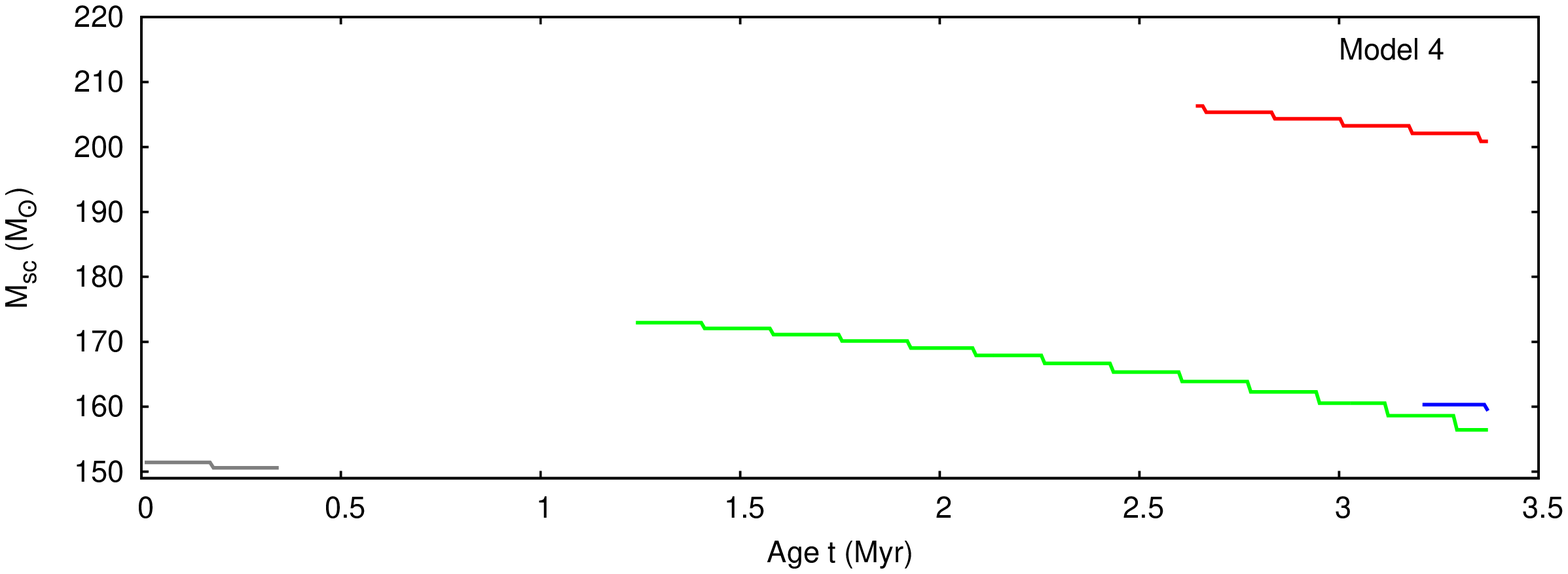}
\vspace{-0.5cm}
\caption{Formation of SC stars in the computed ``Model 3'' (upper panel) and ``Model 4'' (lower panel) of \citet{bko2011}.
The panels show the masses of the SC stars at the times of their first appearance and their subsequent mass depletion due to
stellar wind as obtained from within NBODY6. For ``Model 3'' (upper panel), the SC stars represented by the grey
lines, that appear from the very beginning,
are the ``spurious'' ones in this example in the sense explained in Sec.~\ref{xtracomp}. Similar description applies
for the lower panel.}
\label{fig:sc3}
\end{figure*}

\citet{wk2004} for the first time gave a critical look at the question of the existence of a physical upper-limit
of $\approx 150\Ms$ in the IMF, as observed in R136 (the central massive star cluster in the 30 Doradus region of the LMC).
Assuming R136 has a mass in the range $5\times 10^4 < M_{cl} < 2.5\times 10^5\Ms$ \citep{sel99}, they found that
for a canonical IMF with $\imfmax=\infty$, $10 < N(>150\Ms) < 40$ stars are missing in R136 that have masses $>150\Ms$.
The probability that no stars are observed among the 10 expected ones, assuming $\imfmax=\infty$, is $p=4.5\times10^{-5}$,
\ie, the observed massive stellar content of R136 implies a physical stellar mass limit at $\imfmax\approx150\Ms$.
Similarly, \citet{fig2005} found a dearth of $N(>150\Ms)=33$ stars in the Arches cluster, where the shallowing
of its stellar mass function due to its rapid tidal dissolution has been incorporated. The corresponding number of missing
stars for a canonical IMF is $N(>150\Ms)=18$ which gives a non-detection probability as small as $p=10^{-8}$. Thus the
Arches cluster also has $\imfmax\approx150\Ms$ with a very high significance. Finally, \citet{oey2005} studied 9 clusters
and associations in the Milky Way, LMC and SMC to investigate the expected masses of the most massive stars in these
for different upper mass limits (120, 150, 200, 1000 and $10000\Ms$). They concluded that the observed number
of massive stars supports the existence of a general physical upper mass cutoff within the range
$120\Ms < \imfmax < 200\Ms$ with a high significance.

The massive stellar population of young massive star clusters therefore indicates a physical upper mass limit
near $\imfmax\approx150\Ms$ which is the canonical upper limit of the stellar IMF. One can regard this
mass as the limit imposed by the process of star formation under physical conditions achievable in a
star-forming core. The origin of this very limit is under investigation and the value should
currently be taken as empirical only. As coined by \citet{pk2011}, a stellar population containing stars with their
zero-age-main-sequence (ZAMS) masses up to $150\Ms$ is ``saturated''.

In this paper, our concern is again related to the massive stellar population in R136.
The particular aspect that we focus on is related to a
recent study by \citet{crw2010}. These authors re-analysed the massive stellar population of
R136 in unprecedented detail using Hubble Space Telescope and
Very Large Telescope spectroscopy and high spatial resolution near-IR photometry to find 4 stars,
within the central $1\times1$ pc of R136, with masses $165 - 320\Ms$, \ie, substantially above the canonical limit
One can call such a stellar population, containing single stars with masses substantially exceeding $150\Ms$,
as ``super-saturated'' and the stars surpassing the canonical upper limit can be called
``super-canonical'' (\citealt{pk2011}; hereafter SC). Although the stellar population of R136
has been studied by earlier authors, the crucial turn in \citet{crw2010} is due to these authors' consideration
that the observed SC stars in R136, in spite of exhibiting WN-type spectra and possessing strong winds,
are actually young, main-sequence stars rather than classical Wolf-Rayet (WR) stars. To take this into account,
\citet{crw2010} incorporate detailed stellar evolutionary models of very massive main-sequence stars
that include state-of-the-art treatment of stellar winds which lead to the inference of the SC masses.

More recently however, \citet{bko2011} (hereafter Paper I) showed that single stars with masses $m_s>150\Ms$ can indeed form in
R136 through mergers of the members in massive binaries. These authors computed the evolution of
model clusters that resemble R136 using the state-of-the-art direct N-body integration code ``NBODY6''
\citep{ar2003}. In these computationally challenging models, all the O-stars were taken to be in tight binaries as observations
indicate (\citealt{sev2010}; see Sec.~\ref{compute}) but the ZAMS mass of a single component never exceeds $\imfmax=150\Ms$.
They found that in course of their dynamical evolution, the massive binaries in the model clusters can often merge
due to hardening and/or eccentricity enhancement due to dynamical encounters to produce
single stars of $m_s>150\Ms$. This implies that the presence of SC stars in R136
does not illustrate a violation of the canonical upper limit.

Of course, the value of the upper stellar mass limit is not necessarily exactly $=150\Ms$
\citep{oey2005}. However, for definiteness, we take the following \emph{hypothesis}: there exists a fundamental
upper mass limit of $\imfmax=150\Ms$ or the canonical limit of ZAMS stars formed in clusters  
and any observed more massive (``super-canonical'') star is created from
dynamically-induced stellar mergers within the dense young cluster.
In this study, we aim to test the above \emph{hypothesis} or answer the following question:
\emph{Can the observed number of super-canonical stars in R136 be explained by the naturally
occurring stellar-dynamical and stellar-evolution processes in R136?} To that end, we utilize
the 4 direct N-body computed massive cluster models of Paper I mimicking R136
(Sec.~\ref{initcond} \& \ref{evolmethod})
and additionally perform 5 more similar computations (Sec.~\ref{xtracomp}) to trace the formation of SC stars
through dynamical means. These computed models initiate with properties that are consistent with the
observed properties of young clusters, in particular,
primordial mass segregation \citep{lit2003,chen2007} and massive stellar binaries
with component mass ratio close to unity \citep{sev2010}.
Furthermore, we utilize detailed rotating stellar evolution models in the SC mass range to
estimate the mass evolution of the dynamically formed SC stars and the resulting lifetimes
in their SC phases (Sec.~\ref{scwind}). We conclude the paper by summarizing our results and pointing out the limitations
of the present study (Sec.~\ref{discuss}).

\section{Computations}\label{compute}

\begin{figure*}
\centering
\includegraphics[height=6.0cm,angle=0]{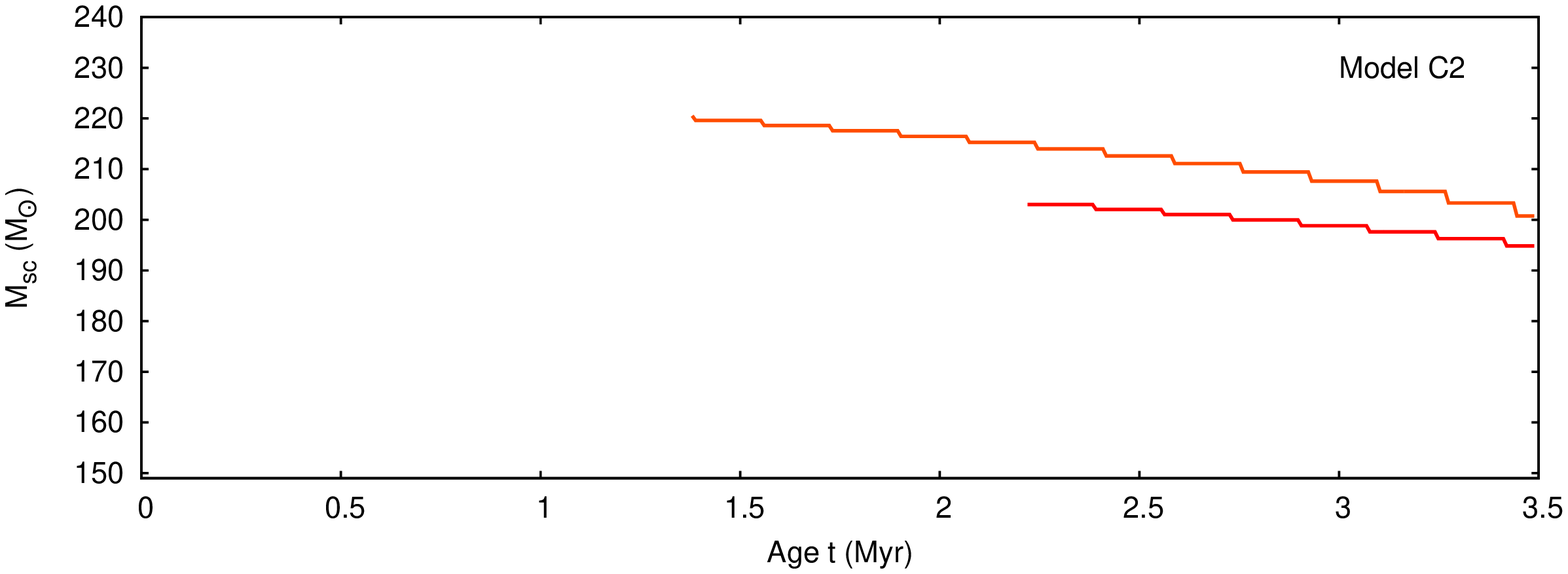}\\
\vspace{-0.6cm}
\includegraphics[height=6.0cm,angle=0]{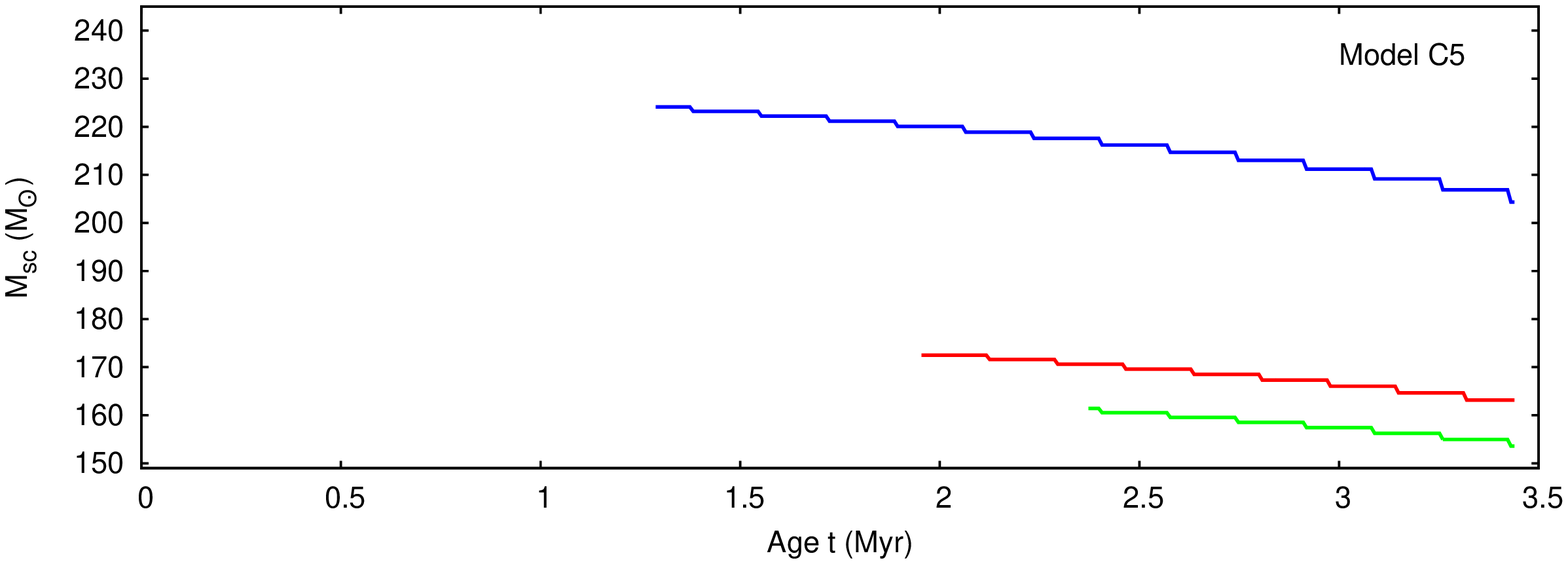}\\
\vspace{-0.6cm}
\includegraphics[height=6.0cm,angle=0]{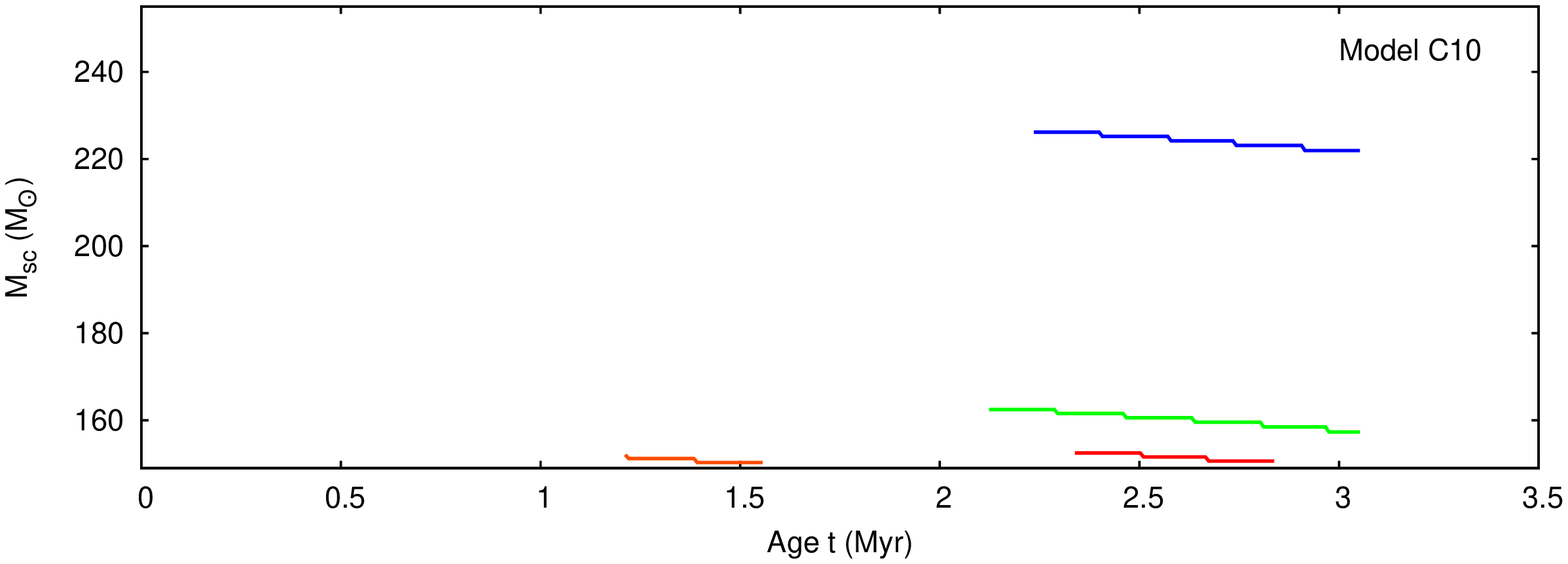}
\vspace{-0.5cm}
\caption{Formation of SC stars in the models ``C2'', ``C5'' and ``C10'' with all their primordial binaries initially in
circular orbits (see Sec.~\ref{xtracomp}). Unlike the computations in Paper I (\cf Fig.~\ref{fig:sc3}),
there are no ``spurious'' SC stars in these cases.}
\label{fig:circbin}
\end{figure*}

State-of-the-art calculations have been performed in Paper I to evolve model star clusters
that mimic R136 in terms of observable parameters. The primary objective of these calculations was to
study the ejected OB-stars from R136, in particular, whether massive runaway stars like VFTS 682 and 30 Dor 016
can indeed be ejected dynamically from R136 as suspected \citep{evns2010,blh2011}.
Besides good agreement of the kinematics of the massive ejecta with those of these noted runaways,
it is also noted in Paper I that the tight massive binaries which drive these runaways can merge due to
the dynamical interactions leading to the formation of SC stars in each of the models, in agreement
with the observed stellar population in R136 \citep{crw2010}.
In this study, we continue to utilize the results of these computations.
The initial conditions and the method of these calculations are accounted in detail in Sec.~2 of Paper I which
we briefly recapitulate here.

\subsection{Initial conditions}\label{initcond}

The above computations comprise direct N-body integrations of
Plummer spheres \citep{pk2008} with parameters conforming with those of R136. The initial mass
of the Plummer spheres is $M_{cl}(0) \approx 10^5\Ms$ which is an upper limit for R136 \citep{crw2010} and
the half mass radii are taken to be $r_h(0)\approx0.8$ pc. The clusters are made of stars with ZAMS masses
drawn from a canonical IMF \citep{pk2001} over the range $0.08\Ms < m_s < 150\Ms$ and are of metallicity
$Z=0.5Z_\odot$.

As for the primordial binary population, stars with $m_s > 5\Ms$ are all in binaries while all lighter stars
are kept single. This termination of the binary population is due to computational ease; binaries bottleneck
the calculation speed of direct N-body integration significantly so that adopting a full spectrum of primordial
binaries (\ie, 100\% initial binary fraction) in models of the size that we compute becomes prohibitive. 
Disregarding the binary population for $m_s<5\Ms$ of course does not affect the ejection of massive stars and
mergers of massive binaries significantly (see Paper I for details). Following the observed period
distribution of O-star binaries \citep{sev2010}, the orbital periods of the binaries having primary masses
$m_s>20\Ms$ are chosen from a uniform distribution between $0.5<\log_{10}P<4$, where $P$ is the orbital period in
days. For primary masses $5\Ms<m_s<20\Ms$, the ``birth period distribution'' of \citet{pk95b} is adopted. 
The eccentricity distribution is chosen to be thermal \citep{spz,pk2008}. These distributions, however, are
not modified via ``eigenevolution'' \citep{pk95b} as the binary-systems' internal
interactions between massive pre-main-sequence stars are not yet quantified.

The initial configurations are fully mass-segregated \citep{bg2008} so that the massive binaries concentrate
within the clusters' central region. In other words, it is assumed that the clusters are primordially
segregated which was the case for several Galactic globular clusters \citep{mrk2010}
and young star clusters \citep{lit2003,chen2007}. In these mass-segregated initial models, the stars are re-distributed
over the cluster in such a way that the individual mass groups are in equipartition whilst the overall cluster
is in virial equilibrium as well. The algorithm for creating such systems is described in the appendix of
\citet{bg2008}. We emphasize that we are conforming with
the \emph{observed} states of massive, young clusters by starting our models binary-rich and mass-segregated.
That is, we \emph{are not} choosing some particular fine-tuned initial conditions in order to enhance
our results.

\subsection{N-body integration, stellar evolution and stellar mergers}\label{evolmethod}

\begin{figure*}
\centering
\includegraphics[height=6.0cm,angle=0]{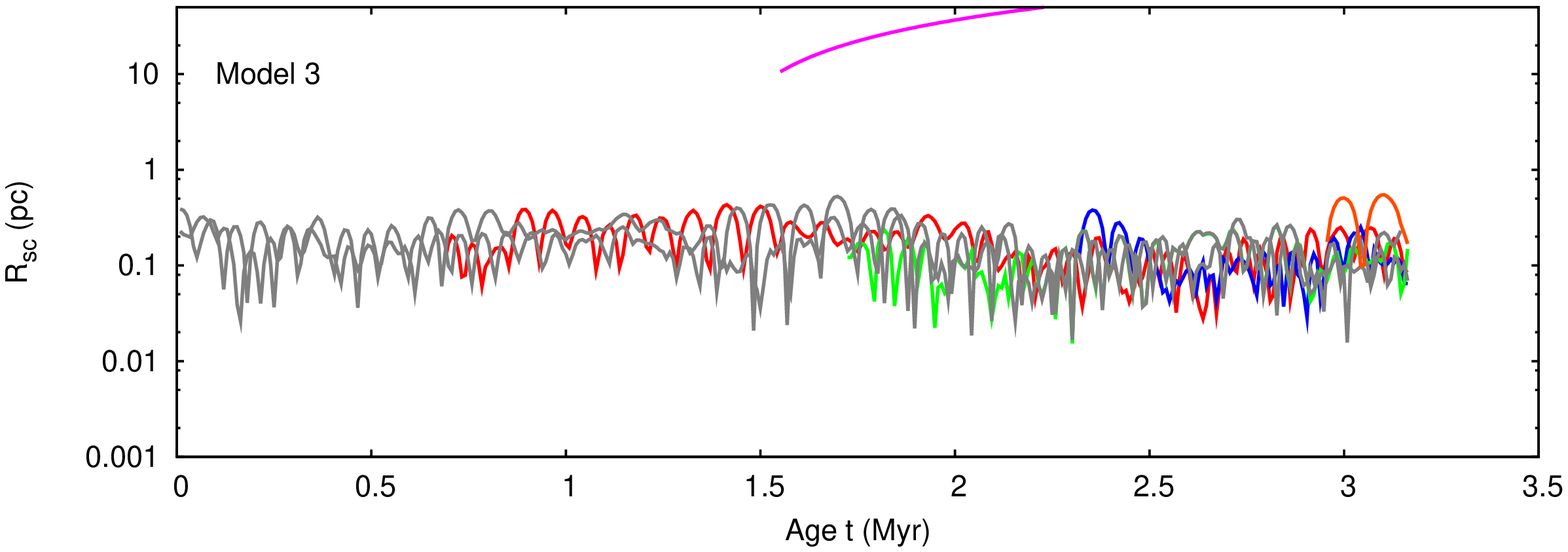}\\
\vspace{-0.6cm}
\includegraphics[height=6.0cm,angle=0]{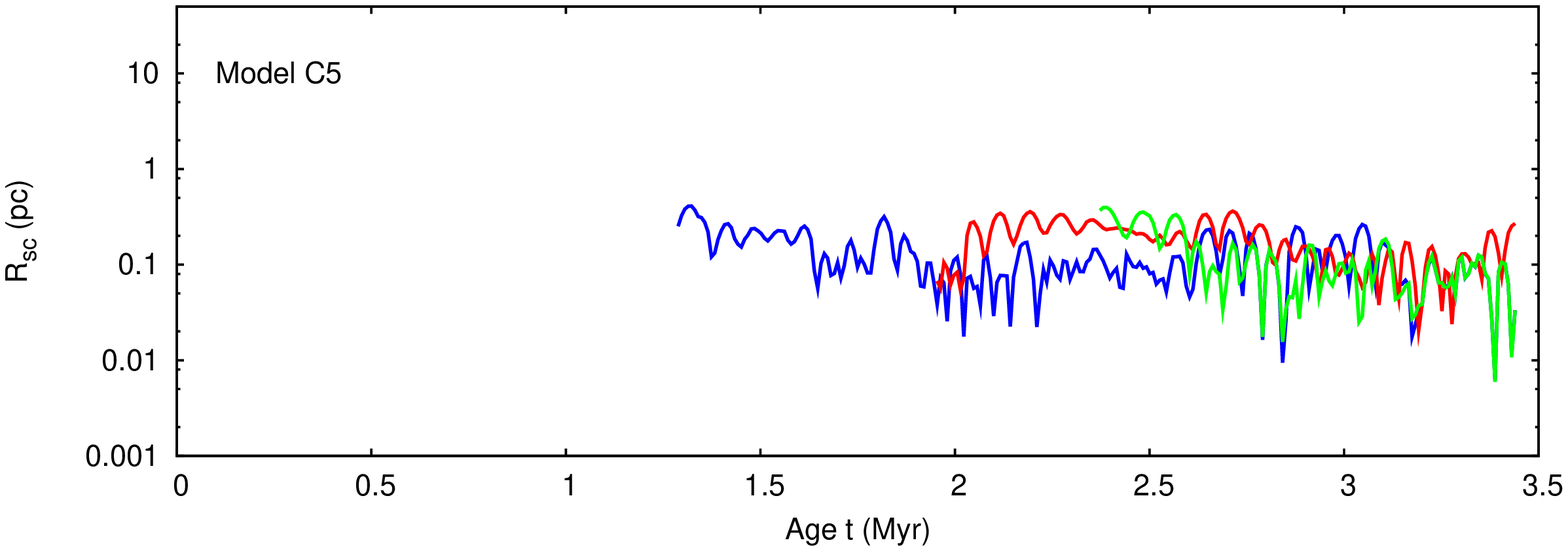}\\
\vspace{-0.6cm}
\includegraphics[height=6.0cm,angle=0]{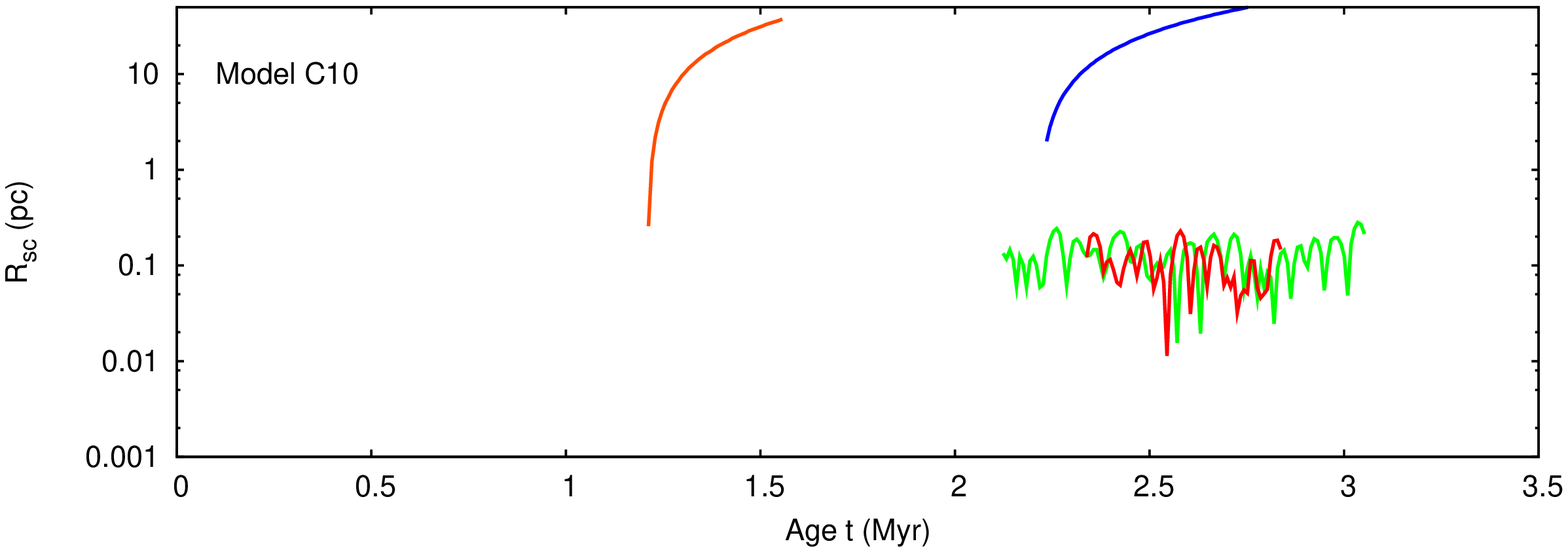}
\vspace{-0.5cm}
\caption{Radial distances, $R_{\rm SC}$, of the SC stars w.r.t. the cluster's center of density that are formed in the models
``3'', ``C5'' and ``C10''. It can be seen that while the SC stars generally form and remain close to
the cluster's center, some of them form with runaway velocities (the monotonically outgoing trajectories; see Sec.~\ref{scform})
and they soon escape as runaway stars from the cluster. The colours of the lines
correspond to the same stars as in Figs.~\ref{fig:sc3} \& \ref{fig:circbin}.}
\label{fig:scrad}
\end{figure*}

With such an initial setup, these computations are the very first direct N-body calculations under such extreme
conditions (all massive stars in binaries, fully mass-segregated). To perform the direct N-body computations,
the state-of-the-art ``NBODY6'' integrator \citep{ar2003} has been used. In addition to integrating the particle
orbits using the highly accurate fourth-order Hermite scheme and dealing with the diverging gravitational
forces in close encounters through regularizations, NBODY6 also employs the well-tested analytical stellar and binary
evolution recipes of \citet{hur2000,hur2002} (the {\tt SSE} and the {\tt BSE} schemes).
These prescriptions are based on model stellar evolution tracks
computed by \citet{pols98}. The wind mass loss of the massive main-sequence (hereafter MS) stars is adopted using the empirical
\citet{nj90} formula. The mass loss rates are, of course, appropriately modified on the giant branches and other
evolved phases for stars of all masses \citep{hur2000}. Admittedly, the \citet{nj90} mass loss rate is too simplistic
for massive stars; it is likely to grossly underestimate the wind mass loss of massive stellar merger products, which
we focus on here, as these objects usually become Wolf-Rayet (WR) stars even in their main sequence phase \citep{gleb2009}.
The above stellar wind recipe is currently the best available one in a direct N-body evolution code; other widely-used direct
N-body codes, \eg, the ``STARLAB'' \citep{pz2001} and the ``NBODY6++'' \citep{spur99} utilize analogous wind treatments. 
The wind mass loss being crucial for very massive stellar entities \citep{gleb2009},
we consider the effects of more detailed stellar winds separately in Sec.~\ref{scwind}.

Mergers among MS stars are treated in NBODY6 by adopting the
schemes of \citet{hur2002,hur2005}.
When two MS stars collide (a collision between two single MS stars or between two MS components
in a binary due to eccentricity induced by close encounters and/or due to encounter hardening),
the two main assumptions in the above scheme are (i) the merged object is a MS star with the
stellar material completely mixed and (ii) no mass is lost from the system during the hydrodynamical process
leading to the merger. This no mass-loss assumption is based on the results of SPH simulations of MS-MS
mergers (\eg, \citealt{sil2001} who show that the mass loss is up to 5\% for low-mass MS stars)
but such studies also show that the mixing is only partial. The
age of the merged MS star is assigned based on the amount of unburnt hydrogen
fuel gained by the hydrogen-burning core as a result of the mixing. In case of a
mass transfer across a MS-MS binary, the cases of the original accretor MS star
having a radiative and a convective core are treated separately (included in the {\tt BSE} algorithm).
For a convective core, the core grows
with the gain of mass and mixes with the unburnt hydrogen fuel so that the accreting MS
star appears younger. For the case of a radiative core, the fraction of the hydrogen burnt
in the hydrogen-burning core remains nearly unaffected by the gain of mass
so that the effective age of the MS star decreases. However, in any case of a merger between two MS
stars, an appropriate amount of mass is removed from the final product if the kinetic energy of their
final approach exceeds the binding energy of the merged MS star.
Admittedly, there exists some arbitrariness in applying the treatments, appropriate for mergers
of low-mass stars, to the cases involving mergers of $\approx100\Ms$ stars which we consider in the following
sections. The structure of these
stars is very different so that it is unclear whether the above collisional mass-loss and mixing
criteria will remain valid for such massive stellar mergers. Unfortunately, the outcomes of massive stellar
collisions are currently unclear from theoretical studies. A relatively simplistic but very fast algorithm
for treating collisions of massive stars (the {\tt MMAMS} scheme) by \citet{gabu2008} shows $<10\%$
mass loss and partial mixing in encounters of $\sim 10\Ms$ stars. In the absence of a clear understanding,
we continue to utilize the scheme for low mass stellar mergers for massive mergers also.

Note that the computed Models 1-4 of Paper I constitute exactly the same cluster model of 4
different randomized (discrete) realizations. Each of these models are evolved upto $\approx 3$ Myr
which is the widely used age of R136 as inferred from its low mass stellar population \citep{and2009}.
The above age estimate is perhaps the most robust one for the R136's stellar population.

\subsection{Additional model computations}\label{xtracomp}

In the four model calculations presented in Paper I (Models 1-4), there are typically 1-2 massive binary mergers
per cluster at the very beginning of the computations that form SC stars. These mergers are found to happen
mostly due to high initial eccentricities of the corresponding massive primordial binaries. Although these merger
products are highly super-canonical, we consider such immediate mergers as ``accidental''
since the parent binaries would have already
merged during their pre-main-sequence evolution. However, we adhere to the canonical upper limit of $\imfmax=150\Ms$  
for R136 which is a plausible assumption as the limit is deduced from observations of young massive star clusters.
Hence, we consider the SC stars, that appear due to a collision between the two members of a primordial binary at the very
beginning of the cluster evolution (\cf Fig.~\ref{fig:sc3}), as spurious and unrelated to the formation of the
initial ZAMS stellar population. These are artifacts of the chosen eccentricity distribution.
The SC stars that are formed later in the course of the cluster evolution are, of course, considered genuine.

Notably, the presence of 1-2 spurious (single) SC stars is not expected to affect the dynamical evolution of our
model clusters in any significant way as for a massive primordially mass-segregated system like ours,
there are anyway a large number of massive stars in the cluster's central region. Therefore the inclusion of
a few more massive single stars is unlikely to provide a substantial additional effect. It is, of course, not
impossible that two massive stars happen to collide to form a single SC star at the birth of a cluster, but
it seems unphysical that it could be an eccentricity effect which is always the case for the above mentioned accidental
SC stars, which is why we consider them spurious. The conditions under which such ``primordial-mergers'' can happen
and their chances is beyond the scope of the present study. In any case, apart form their
negligible contribution to the dynamics of their host clusters, it is unlikely to find any such primordially-merged
star in its SC state in R136 at its current age, due to the short SC lifetime,
so that the currently observed ones should all be dynamically formed. We elaborate the latter point in Sec.~\ref{scobs}.

To see whether the chosen thermal eccentricity distribution in the computed models of Paper I
does have a systematic effect on the dynamical formation of SC stars,
we perform 5 additional computations with the same initial configuration (but different realizations)
as in Paper I (see Sec.~\ref{initcond}) except that all the primordial binaries are initially taken to be circular.
Of course, the most realistic
approach would be to ``eigenevolve'' \citep{pk95b}, \ie, to incorporate the pre-main-sequence evolution of each of
the primordial binaries to determine their eccentricities and mass-ratios at the beginning of the
computations. However, the nature of eigenevolution is not yet known for progenitors of massive stars.
Typically, the tidal interaction between the proto-massive-star members
would circularize the tight binary orbits but wide binaries would still remain eccentric. Currently,
it is unclear in the literature what would be the exact outcome of eigenevolution of massive proto-stellar
binaries. As we shall see in Sec.~\ref{res}, if only the dynamically formed SC stars are considered,
their population is similar for the calculations in Paper I and in these new computations, \ie, the primordial
binaries' eccentricity distribution does not play a significant role.

We discuss the formation of SC stars from the computations of Paper I and from these new
computations in Sec.~\ref{res}.

\section{Results: formation of super-canonical stars}\label{res}

Fig.~\ref{fig:sc3}
(top panel) shows the masses of the SC stars, $M_{\rm SC}$, with the cluster evolution time for the
computed ``Model 3'' of \citet{bko2011}. Each appearance of a line in this
figure corresponds to the appearance of a SC star and its slow decline is due to the wind mass loss of the SC star
as provided by the currently available stellar wind prescriptions in NBODY6 (see Sec.~\ref{evolmethod}).
As pointed out in Sec.~\ref{evolmethod}, this stellar wind scheme is inappropriate for very massive stars like the
super-canonical ones, as it substantially underestimates the mass loss and enhances the lifetime
in the SC phase. We address this issue in Sec.~\ref{scwind}.

The SC stars represented by the grey lines that appear from the very beginning
of the evolution are formed due to immediate mergers of highly eccentric massive primordial binaries which are
unintended and considered spurious as explained in Sec.~\ref{xtracomp}. The rest of them are of course genuine
SC stars that are formed through dynamically-induced mergers of massive binaries (see Sec.~\ref{scform}).
This particular model formed a rather marked number of SC stars. Notably, a $M_{\rm SC}\approx 240\Ms$ SC star is formed
for the first time as early as $T_0=0.7$ Myr.

Fig.~\ref{fig:sc3} (bottom panel) similarly shows the formation of SC stars for ``Model 4'' of Paper I.
Here, only one spurious SC star is
formed which is only marginally more massive than the canonical upper limit
and hence disappears soon due to wind mass loss. The subsequent
cluster contains only dynamically formed SC stars. In this model also, the first appearance of a
SC star ($M_{\rm SC}\approx 170\Ms$) occurs quite early ($T_0=1.2$ Myr).

Fig.~\ref{fig:circbin} shows the formation of SC stars in the additional models ``C2'', ``C5'' and ``C10'' with
all primordial binaries initially circular as explained in Sec.~\ref{xtracomp}. While the total number of
SC stars formed in these models are typically similar to those in the models of Paper I (\cf Table~\ref{tab1}),
none of them is formed artificially as can be seen in Fig.~\ref{fig:circbin}. The SC stars start to appear from
typically $T_0\approx 1$ Myr for these models also. The remaining two of the additionally computed models
produce SC stars only with masses marginally above $150\Ms$. 

\subsection{Dynamical formation of SC stars}\label{scform}

In the above computations, we find several dynamical channels for the formation of the SC stars.
The most frequent way is found to be an abrupt eccentricity enhancement of a massive binary due to a close
encounter with a single star. A collision and subsequent merger occurs between the components
when their periastron separation becomes smaller than the sum of their radii.
The single stars in the cluster's central region arise from dissociation
of wider binaries by encountering with harder binaries or by direct ionization (\ie, detachment
of a binary due to its encounter with a single star of kinetic energy larger than its binding energy).
This channel is most
fruitful for the Models 1-4 as they generally contain more eccentric binaries.

The next important channel is the eccentricity augmentation of the inner massive binary due to Kozai
cycles with an outer companion. Such hierarchical triples can often form in the cluster center
as a result of binary-binary encounters. Notably, a close encounter of a single star with a hard binary
can also form a temporary triple system (resonance encounters; see \citealt{hh2003}). The orbital changes due to
tidal interactions within the inner massive binary and also that of the outer member with the inner members
and as well the merger process of the inner binary can often make the triple unstable. This causes the inner
binary to become isolated from the outer companion in the course of its eccentricity growth. In our computations,
a few newly formed SC stars are still found bound in wide orbits with the outer member
which soon get ionized (within a few orbital times).
In a few cases, a SC star is also found to form when the outer member of a triple collided with one of the
inner members.

SC stars also form due to encounter hardening \citep{hg75} of the orbits of appropriate binaries resulting
in a Roche lobe overflow. The approach to the Roche lobe contact is generally assisted by the evolutionary expansion
(during its main sequence) of the more massive binary member which therefore fills the Roche lobe first (\cf \citealt{hur2005}).
The subsequent unstable mass transfer from the more massive member to the less massive one coalesces
the binary.

Some of the SC stars are found at high velocities at their birth. These are the ones whose progenitor binaries
suffered substantial
recoils in the close encounters that caused them to merge and they escape from the cluster
right at their birth. As examples, the radial distance, $R_{\rm SC}$, vs. $t$ plots for the SC
stars in Fig.~\ref{fig:scrad} show such runaways in the models ``3''and ``C10'' exhibiting 
the monotonic outgoing trajectories.
In fact, such born escapers constitute most of the runaway stars in the SC mass-range. 
This can be expected since once formed, it is unlikely to eject a SC star later by a super-elastic close encounter
\citep{hg75} with a hard primordial binary unlike the non-SC stars for which such an ejection channel is
usual. The SC star, being the most massive participant in such an encounter (the primordial binary
members are $<150\Ms$), is likely to get trapped in the binary by exchanging with one
of the companions and hence receiving less recoil.

The rest of the SC stars initially form and remain 
bound to the cluster. Because they are among the most massive members of the cluster, they continue to reside
close to the cluster's center, occasionally several of them simultaneously (\cf Fig.~\ref{fig:scrad}).
However, a few of them can subsequently be ejected by dynamical recoils (only two such SC ejecta occurred
from all of our computations). Notably, the ejection or runaway fraction
of stars in the SC mass range is $>0.5$ as determined in Paper I (see Fig.~4 of Paper I).

It is to be noted that the above dynamical channels are as well applicable for formation of merger products
with $m_s<150\Ms$. 

\section{Evolution of super-canonical stars: the effect of stellar winds}\label{scwind}

\begin{figure*}
\centering
\includegraphics[width=15.0cm,angle=0]{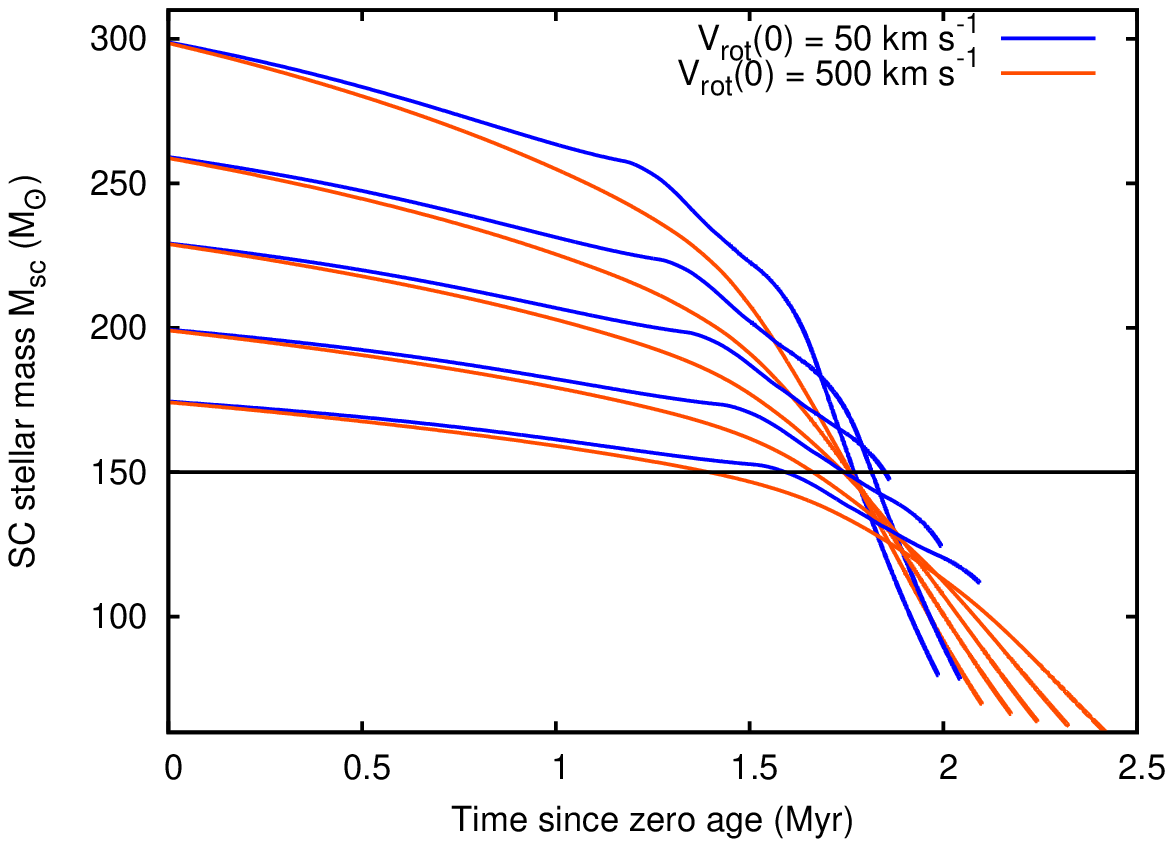}
\caption{Evolution of the mass of initially SC stars as computed by \citet{kohl2012} for initial surface rotation velocities
of $V_{\rm rot}(0)=50$ \& $500 {\rm ~km ~s}^{-1}$ and metallicity appropriate for the LMC.
Followed by an initial moderate mass loss,
as given by the \citet{vink2001} stellar wind, a SC star undergoes a substantial mass loss as it enters the WR phase.
The black horizontal line highlights the canonical upper limit of $150\Ms$. It can be seen that the stellar mass remains
super-canonical until $\approx1.4-1.8$ Myr from the ZAMS. See text for details.}
\label{fig:scevol}
\end{figure*}

As already raised in Sec.~\ref{initcond}, the treatment of stellar wind mass loss in NBODY6 is inappropriate
for massive (main-sequence) entities like the SC stars which grossly underestimates their wind mass loss, thereby
enhances their lives in the $m_s>150\Ms$ state. In Fig.~\ref{fig:sc3}, it can be seen that all the SC stars formed
in ``Model 3'' remain super-canonical until the end of the computation at $t\approx 3$ Myr. This is as well true
for all the substantially super-canonical merger products in the other computations.

To take into account an appropriate mass-loss for the SC stars, we consider the stellar evolutionary
models of rotating MS stars by \citet{kohl2012} which encompass
the SC mass range and are computed for the LMC metallicity. The above
authors use the one-dimensional hydrodynamic binary evolution code of \citet{heg2000} which includes direct treatments
of stellar rotation, magnetic field and detailed stellar winds. The code has been recently improved as in
\citet{petv2005} and \citet{yoon2006}.

Stellar wind is the most important factor for the evolution of the mass of an isolated massive star during its main
sequence. In the computations by \citet{kohl2012}, the MS stellar wind of \citet{vink2001} is applied that includes
the bi-stability jump. During the bi-stability jump,
a linear interpolation is applied as in \citet{brot2011a}. The mass loss rate is switched from \citet{vink2001} to
that of \citet{nj90} when the latter becomes stronger (at effective temperatures $<22000$ K). When the surface Helium
abundance becomes $Y_s \geq 0.7$, the star is considered to enter the WR phase in which case
the much enhanced \citet{hama95} wind, divided by a factor of 10 \citep{yoon2006}, is applied.

Fig.~\ref{fig:scevol} shows the resulting mass evolution of SC stars with initial surface rotation velocities of
$V_{\rm rot}(0)=50$ \& $500 {\rm ~km ~s}^{-1}$. The SC stars undergo a steep mass loss as they become Wolf-Rayets.  
Because of the large mass loss in the WR phase, these stars become less massive than $150\Ms$,
\ie, cease to be super-canonical, in $\tsc\approx 1.4 - 1.8$ Myr from their zero-age. The lifetime in the SC phase
therefore becomes much shorter with the more detailed mass loss prescription in a hydrodynamical stellar evolution 
calculation. We consider this more realistic lifetime as the lifetime of a SC star rather than that obtained
from within NBODY6. We note that the SC stars formed in our computations do not undergo further coalescence
or Roche lobe overflow with a secondary (although these are in principle possible) so that the individually obtained
mass loss rates can be applied to them (see also Sec.~\ref{discuss}). It can be noted from Fig.~\ref{fig:scevol}
that the mass loss rate is only moderately affected by the stellar spin. The SC stars being binary merger products
are indeed likely to form with high surface rotation velocities. 

Notably, in a dynamically active environment, a SC star can undergo
further mass evolution due to subsequent merger of the SC product with other stars or its acquirement
of a binary companion causing Roche lobe overflow. However, in our computations, we hardly find
subsequent mergers of the SC stars or them being overflown as confirmed by their continuous and
smooth mass depletion (\cf Figs.~\ref{fig:sc3} \& \ref{fig:circbin}), justifying
the application of isolated stellar mass loss to them.

\subsection{Can the dynamically formed SC stars in R136 be observed?}\label{scobs}

Given that the lifetime of the SC phase (for the LMC metallicity) is
$\tsc\approx 1.4 - 1.8$ Myr, how likely is it to detect the dynamically formed SC stars
in R136 at the present day? Clearly, the SC stars must form at most a $\tsc$ time earlier to be presently visible in their
SC phases. From Figs.~\ref{fig:sc3}, \ref{fig:circbin}, \ref{fig:scrad} and Table~\ref{tab1}
(see below), it is clear that  
(a) SC stars are very likely to form at any time within 1 - 3 Myr cluster age and (b) multiple SC stars are likely
to appear and remain bound to the cluster within a SC lifetime of $\tsc=1.5$ Myr,
the latter value being taken representatively. 

These conclusions imply that \emph{in spite of the severe wind mass loss of the SC stars, it is still
feasible that they can be detected at the present day in multiple numbers near the center
of a R136-type cluster}. Such SC stars are the ones that are formed via dynamical mergers of massive
binaries late enough that they remain super-canonical until now. Our calculations indicate that 
the satisfaction of this condition is quite feasible for a R136-type cluster. For example, at the cluster age of 2.5 Myr,
``Model 3'' would retain 3 SC stars (\cf Figs.~\ref{fig:sc3} and \ref{fig:scrad}), C5 and C10 
would retain 3 and 2 SC stars
respectively (\cf Figs.~\ref{fig:circbin} and \ref{fig:scrad}).

By the same argument, \emph{the above detailed evolutionary models of the SC stars
imply that those observed in R136 would have evolved below the canonical
limit by the present day had they been primordial, \ie, had they been formed at the cluster's birth, thereby 
violating the canonical upper limit, assuming the widely used $\approx 3$ Myr age of R136 \citep{and2009}.
This strongly
supports our dynamical formation scenario of the observed SC stars in R136} which causes them to form sufficiently
later so as to make them visible during their SC phases at the present day. This is well supported by our computations.

Admittedly, there are several drawbacks and incompleteness in the above line of arguments. First, the detailed
stellar evolutionary models of \citet{kohl2012} have been considered independently of the dynamics of the model
clusters. The newer stellar evolutionary models would result in stronger wind mass loss from the cluster core
and would contribute to the dilution of the cluster's central potential to a larger extent than that from the NBODY6's
native stellar evolutionary scheme \citep{hur2000,hur2002}. However, in our computations the cluster's
core, where the binary mergers occur, is populated by a large number of hard massive binaries due to our
initial conditions (Sec.~\ref{initcond}), \emph{which are all in accordance with observations
of young star clusters} (see Sec.~\ref{initcond} and references therein). In that case, the dynamics of the
cluster's central region will be dominated by the energy generated in the super-elastic binary-single and
binary-binary encounters \citep{hg75,hh2003} and the stellar mass loss would not affect the 
dynamical encounter rates in the cluster's center significantly.

It is true that the mass loss from the individual
companions of a binary would result in expansion of the binary orbit and hence dilute its hardness. However,
as shown by \citet{fpz2011}, this effect is counteracted by the dynamical hardening \citep{hg75} of the binaries so that they
remain hard throughout. In the N-body calculations of \citet{fpz2011}, even just a single binary survives
to remain hard in their model cluster's core and dominates the energy generation in the cluster's core
(until its ejection). With a large number of binaries, as in our case, the encounter hardening and the
corresponding energy generation would be substantially stronger as the binaries provide much larger encounter cross
sections than single stars due to their much bigger geometrical size (see also \citealt{fuj2012}).
Noticeably, even after including stellar
wind mass loss, the computed single-star-only models of \citet{fpz2011} undergo deep core-collapse within $\approx 3$ Myr,
\ie, the effects of dynamical relaxation and negative specific heat \citep{spz} themselves dominate over the wind
mass loss. Our models, of course, do not core collapse due to the substantial energy release from the central primordial
binaries.

Another drawback of the above arguments is that they are based on stellar evolution models that begin
from ZAMS with normal composition. The SC stars are merger products of O-stars that are evolved from their
zero age and therefore would be He-enriched and substantially rotating. While we do consider rotating models,
the initial He fraction is taken to be that appropriate for the LMC, \viz, $Y=0.2562$ \citep{kohl2012}. To
obtain a basic estimate of the He-richness of the merged product, let us consider two $100\Ms$ stars that merge to
form a typical $200\Ms$ SC star at a typical cluster age of 1.5 Myr. Each of the initial stars (before merger) would
possess a H-burning core of $M_c\approx80\Ms$ and at the above merging age their core He fraction 
$Y_c\approx0.5$ (\ie, $\approx40\Ms$ He per star) according to the above newer models
(Karen K\"ohler; private communication). Assuming an ideal complete mixing and no mass loss during the merger process,
the new He abundance of the merged product would be $Y\approx0.4$ which would also be the rejuvenated
surface value $Y_s$; for a partial mixing the surface abundance would be lower. While this $Y_s$ is
large enough to drive a stronger wind from the newly formed SC star compared to that for its ZAMS
value $Y_s\approx0.25$, it is still moderately above the ZAMS value and less than the abundance at which
the WR winds turns on ($Y_s\geq0.7$; see Sec.~\ref{scwind}). Therefore, a drastically large wind from the
newly formed SC star is not expected.

The above discussions imply that none of the shortcomings or uncertainties of our approach dictate to the
impossibility of observations of SC stars in R136; rather such observations are still feasible.

\section{Discussion and Summary}\label{discuss}

\begin{table*}
\centering
\begin{minipage}{138mm}
\centering
{\large
\caption{\large Table summarizing the formation of SC stars in our computed models. The descriptions of the columns
are as follows: Col.~(1): model ID, Col.~(2): time $T_0$ at which a SC star first appeared in the model, Col.~(3): birth mass
$M_0$ of the first-comer SC star, Col.~(4): birth mass $M_{\rm max}$ of the most massive SC star formed in the model,
Col.~(5): formation time, $T_{\rm max}$, of the most massive SC star, Col.~(6): maximum number of SC stars,
$\mathcal{N}_{\rm max,in}$, that remained simultaneously bound to the cluster over a $\tsc=1.5$ Myr period
within $<3$ Myr cluster age,
Col.~(7): total number of SC stars, $\mathcal{N}_{\rm tot}$, formed over the computation
(all of them do not necessarily appear simultaneously or remain bound to the cluster).
For Models 1-4 (from \citealt{bko2011}) the ``spurious'' SC stars
(see Sec.~\ref{xtracomp}) are excluded and only those SC stars that are formed later in these computations are
considered.}
\label{tab1}
\begin{tabular}{@{}ccccccc}
\hline
Model ID & $T_0$ (Myr) & $M_0$ ($\Ms$) & $M_{\rm max}$ ($\Ms$) & $T_{\rm max}$ (Myr) &
$\mathcal{N}_{\rm max,in}$ & $\mathcal{N}_{\rm tot}$\\
\hline
1 & 2.6 & 193.9 & 193.9 & 2.6 & 1 & 2\\
2 & 2.0 (3.0)$^a$ & 155.2 (181.4)$^a$ & 181.4 & 3.0 & 1 & 2\\
3 & 0.7 & 236.8 & 246.0 & 1.5 & 4 & 5\\ 
4 & 1.2 & 172.5 & 206.2 & 2.6 & 1 & 2\\
C2 & 1.4 & 220.6 & 220.6 & 1.4 & 1 & 2\\
C5 & 1.3 & 224.0 &  224.0 & 1.3 & 3 & 3\\
C10$^b$ & 1.2 (2.1)$^a$ & 152.4 (162.5)$^a$ & 225.9 & 2.2 & 2 & 4\\
\hline
\end{tabular}
\footnotetext[1]{The first-comer SC star's mass is too close to $150\Ms$. The time and the mass corresponding to the next SC
appearance is also shown in the parentheses.}
\footnotetext[2]{The remaining two of the additionally computed models (with initially only circular binaries) produce SC stars
only with masses marginally above $150\Ms$.}
}
\end{minipage}
\end{table*}

Table~\ref{tab1} summarizes the SC star formation in the computed models of Paper I and in the new computations.
Our general conclusions from the computations are as follows:
\begin{itemize}

\item Formation of SC stars due to dynamically induced mergers of massive binaries are common in a
R136-type cluster. In most of the computations we find multiple SC stars formed within $<3$ Myr
which is appropriate for R136 (\citealt{and2009}; see Figs.~\ref{fig:sc3} \& \ref{fig:circbin}).
Most SC stars are formed single while a few of them initially form in wide binaries.

\item The SC stars typically begin to appear from $T_0\approx 1$ Myr cluster age or even earlier
(\cf Table~\ref{tab1}). They tend to form with equal likeliness over a cluster age of 1 - 3 Myr. 

\item The most massive SC star formed in a given computed model is typically close to $M_{\rm max} \approx 200\Ms$ and the
most massive one formed is $\approx 250\Ms$ (``Model 3''; \cf Table~\ref{tab1}). The cluster age $T_{\rm max}$
corresponding to the formation of the most massive SC star is typically well within $<3$ Myr
(\cf Table~\ref{tab1}).

\item Multiple SC stars are occasionally found to exist simultaneously near the cluster's
center which are bound to the cluster, over a representative SC phase lifetime of
$\tsc=1.5$ Myr, within $<3$ Myr (\cf Table~\ref{tab1}) cluster age. This, along with the second point above,
implies that it is quite plausible that R136 harbours multiple SC stars at the present day. 

\item Some SC stars are formed with runaway velocities and escape (\cf Sec.~\ref{scform}; Fig.~\ref{fig:scrad}).
 
\end{itemize}

These conclusions conform with the observations by \citet{crw2010} who found 4 SC stars in R136
in the initial mass range of $165-320\Ms$. In our computations, SC stars of up to $\approx 250\Ms$ are formed
which is consistent with the observations taking into account the large uncertainties in the stellar evolution
models in this mass range. Also, only one of the models (``Model 3'') has up to 4 SC stars simultaneously
bound to the cluster (over a 1.5 Myr period) although other models also host multiple SC stars close to
the cluster's center. Some models however contain only one bound SC star within 
3 Myr (\cf Table~\ref{tab1}).

At this point, it is worthwhile to note that our above conclusions depend somewhat on the age of the R136 cluster
which is, of course, not fully settled yet. While the bulk of R136 is $\approx 3$ Myr old \citep{and2009}
the high-mass stellar population might be younger \citep{mhunt98,dkot98}. In our computations,
the SC stars typically begin to appear from $\approx1$ Myr cluster
age and several of them appear by $\approx 2.5$ Myr. This time-frame is quite consistent with the
possible $\lesssim 2$ Myr age of R136's massive stellar population given the wide uncertainties in the age
of such a young cluster particularly for the massive end of the IMF. There has so far been
no direct estimate of the age of the massive end of the stellar population of R136
and the above age-limit is based only on comparisons with the stellar populations of Car OB1
and NGC 3603 \citep{crw2010}. Given that the stellar IMF of R136 continues to maintain
the canonical law \citep{pk2001} from the low to the high mass range
($1.1\Ms-120\Ms$; \citealt{mhunt98,and2009}), we find it more natural to consider that
the whole stellar population of the R136 cluster has formed in a single starburst event whithout
a significant age spread. The issue can, of course, be more resolved with better understanding of the
evolution of very massive stars and their winds. Finally,
a recent study by \citet{chi2012} shows that the O-star binary distribution can,
in fact, be even harder (\ie, more bound or tighter) than that considered
in our computations (see Sec.~\ref{initcond}). This would lead to the formation of SC stars
even earlier which constitutes an important future study.

It is important to remember that the formation of SC stars at early ages in the above computations
is facilitated by the adopted initial complete mass segregation. This condition subjects the massive
binaries to strong dynamical encounters from the beginning of the cluster evolution. Note however
that primordial mass segregation is inferred to have been true for several Galactic
globular clusters \citep{bg2008,mrk2010,strd2011} and open clusters \citep{lit2003,chen2007}.
Notably, for a completely unsegregated model, it would take $\approx 10$ Myr
for the massive binaries to segregate to the cluster's center which makes the
early formation of SC stars unlikely. However, the mass segregation timescale shortens
substantially with increasing compactness of the cluster. In this context, an important
outlook would be to study the formation of SC stars with varying
initial compactness and degree of primordial mass segregation.

The drawbacks of the computed R136 models, as discussed in Paper I (see Sec.~4 of Paper I),
naturally carry over to the present analyses. These limitations however do not crucially affect
SC formation. In particular, the truncation of the binary distribution at $m_s\approx 5\Ms$ does
not influence the mergers of the most massive binaries, the latter being much more centrally concentrated
due to mass segregation. Also, the exact period distribution of the O-star binaries is not
instrumental for the formation of the SC stars as long as O-stars are largely found in tight binaries
\citep{sev2010}. Notably, our adopted range of the orbital period for the O-star binaries is
similar to that reported by \citet{sev2010} (see Paper I). The formation of SC stars is generally
similar for models with initially thermally distributed eccentricities (Paper I) and for the
newer ones with initially circular binaries (except for the spurious SC stars in the former models).
This implies that the initial eccentricity distribution of the massive binaries does not crucially
influence the formation of SC stars.

The most consistent and accurate way to address the above discussed concerns regarding the effects of a stronger
wind mass loss due to the newer stellar evolutionary models is to incorporate the latter in a direct
N-body code which is well beyond the scope of this paper. No such direct N-body code exists at
the present time. Furthermore, there are substantial uncertainties in the physics of mergers
of massive stars at the present time. Therefore, the conclusions in this work are the best one can 
draw given the current technical limitations. Perhaps it would be possible to do such work in future
with a distributively computing, highly modular N-body calculation framework such as ``MUSE'' \citep{pz2009}.  
In spite of all these uncertainties it is worthwhile to note that if one focuses
to the particular case of R136, then (a) the existence of SC stars is \emph{strongly supported by observations} and (b)
considering the widely used $\approx 3$ Myr age of R136, the SC stars must have formed later than the birth of
R136. Our computations are consistent with the formation of SC stars through dynamically induced
mergers of massive binaries. While there are technologically-limited drawbacks in our present analysis, as
elaborated in Sec.~\ref{scobs}, none of these shortcomings dictate to the non-observation of SC
stars in R136. Therefore, we can say that we have justified the \emph{hypothesis} of Sec.~\ref{intro}
through our detailed N-body modelling of R136 with initial conditions chosen in accordance with observations
of young star clusters. In other words, from our realistically modelled computations of R136-like star clusters,
and from our analyses as presented above, it can now be more definitely concluded that the observed
super-saturated stellar population in R136 does not imply a violation of the canonical stellar upper mass limit
near $\imfmax=150\Ms$.

\section*{Acknowledgments}

We are very thankful to Karen K\"ohler of the Argelander-Institut f\"ur Astronomie, University of Bonn, Germany, for providing us
with the massive stellar evolutionary models and discussing them in detail which have been instrumental for this work. 
We thank Sverre Aarsteh of the Institute of Astronomy, Cambridge, U.K., for his effort in the continuing improvements
of NBODY6 and his help during our computations. We wish to thank Paul Crowther of the University of Sheffield,
U.K., for encouraging discussions. We thank the anonymous referee for guiding criticisms
that led to a substantial improvement of the manuscript.

\label{lastpage}

\end{document}